\documentclass[fleqn,10pt]{wlscirep}
\usepackage[utf8]{inputenc}
\usepackage[T1]{fontenc}
\title{Prediction and Prevention of Pandemics via Graphical Model Inference and Convex Programming}

\usepackage{subcaption}
\usepackage{nameref}
\usepackage{multirow}

\author[1]{Mikhail Krechetov}
\author[2]{Amir Mohammad Esmaieeli Sikaroudi}
\author[2,3]{Alon Efrat}
\author[4]{Valentin Polishchuk}
\author[3,5,2,*]{Michael Chertkov}

\affil[1]{Skolkovo Institute of Science and Technology, 
Moscow, 121205, Russia}
\affil[2]{University of Arizona, Department of Computer Science, Tucson AZ, 85721, USA}
\affil[3]{University of Arizona, Program in Applied Mathematics, Tucson AZ, 85721, USA}
\affil[4]{Link\"oping Univeristy, Communications and Transport Systems, Norrkoping, 60174, Sweden}
\affil[5]{University of Arizona, Department of Mathematics, Tucson AZ, 85721, USA}

\affil[*]{chertkov@arizona.edu}
\begin{abstract}
Hard-to-predict bursts of COVID-19 pandemic revealed significance of statistical modeling which would resolve spatio-temporal correlations over geographical areas, for example spread of the infection over a city with census tract granularity.  In this manuscript, we provide algorithmic answers to the following two {\it \bf inter-related public health challenges of immense social impact which have not been adequately addressed} %by the Applied Math and Theoretical Engineering community}. 
(1) {\bf Inference Challenge:} assuming that there are $N$ census blocks (nodes) in the city, and given an initial infection at any set of nodes, e.g. any $N$ of possible single node infections, any $N(N-1)/2$ of possible two node infections,  etc, what is the probability for a subset of census blocks to become infected by the time the spread of the infection burst is stabilized? (2) {\bf Prevention Challenge:} What is the minimal control action one can take to minimize the infected part of the stabilized state footprint? To answer the challenges, we build a Graphical Model of pandemic of the attractive Ising (pair-wise, binary) type, where each node represents a census tract and each edge factor represents the strength of the pairwise interaction between a pair of nodes,  e.g. representing the inter-node travel, road closure and related, and each local bias/field represents the community level of immunization, acceptance of the social distance and mask wearing practice, etc. {\bf Resolving the Inference Challenge} requires finding the Maximum-A-Posteriory (MAP), i.e. most probable, state of the Ising Model constrained to the set of initially infected nodes. (An infected node is in the $+1$ state and a node which remained safe is in the $-1$ state.) We show that almost all attractive Ising Models on dense graphs result in either of the two possibilities (modes) for the MAP state:  either all nodes which were not infected initially became infected, or all the initially uninfected nodes remain uninfected (susceptible). This bi-modal solution of the Inference Challenge allows us to re-state the {\bf Prevention Challenge} as the following {\bf tractable convex programming}: for  the bare Ising Model  with pair-wise and bias factors representing the system without prevention measures, such that the MAP state is fully infected for at least one of the initial infection patterns, find the closest, {  for example} in $l_1$, {  $l_2$ or any other convexity-preserving} norm, therefore prevention-optimal, set of factors resulting in all the MAP states of the Ising model, with the optimal prevention measures applied, to become safe. {  We have illustrated efficiency of the scheme on a quasi-realistic model of Seattle. Our experiments have also revealed useful features,
such as sparsity of the prevention solution in the case of the $l_1$ norm, and also somehow unexpected features, such as localization of the sparse prevention solution at pair-wise links which are NOT these which are most utilized/traveled.} 
\end{abstract}
\begin{document}

\flushbottom
\maketitle
\thispagestyle{empty}

%\noindent Please note: Abbreviations should be introduced at the first mention in the main text – no abbreviations lists. Suggested structure of main text (not enforced) is provided below.

\section*{Introduction}

We follow our previous work \cite{GM-Pandemic} in justification for the use of the Graphical Models (GM) to study and mitigate pandemics. Therefore, we start from providing a brief recap of the prior literature on modeling of the epidemics, describe the logic which led us in \cite{GM-Pandemic} to the Ising Model (IM) formulation, and then state formally the inference and prevention problems addressed in the manuscript.

Difficulty in both predicting and neutralizing the spread of pandemics is a major social challenge of humanity. Technically speaking, we are yet to design a coherent data lifecycle for modeling and prevention both in terms of the global strategies and local tactics. To address the challenge, we must devise a hierarchy of spatio-temporal models with different resolutions -- from individual to community, county to the city, and from the moment a pathogen first enters our bodies, to days of disease development and to community transmission. Importantly, the models should be efficient in computing probabilistic predictions (for instance, offering the marginal probability heat map for the city neighborhoods to transition from the current/prior state of infection to the projected/a-posteriori state in two weeks).  

Epidemiology and Mathematical Biology experts have relied in the past on a number of modeling approaches. The Agent-Based-Models  (ABMs), introduced in epidemiology in 2004-2008~\cite{eubank_modelling_2004,2005Longini,ferguson_strategies_2005,ferguson_strategies_2006,2006Germann,2008Halloran}, have complemented the earlier  compartmental models~\cite{1910Ross,1927Kermack,1991Anderson,2000Hethcore}. Using ABMs, even though not exclusive to epidemiology \cite{ABM-wikipedia,2018Downey}, became a breakthrough in the field, as they allowed to make a significant improvement in the quality of predictions, especially in the spatio-temporal resolution of how the disease spreads and how one can mitigate its spread. The models became and remained a core part of the epidemiology data life-cycle. (See for instance \cite{ABM-Columbia,ABM-Gates} for most recent bibliography.) The ABMs provide a detailed prediction of how pandemics spread within counties, cities, and regions.  A majority of the country-, city- or county- scale testbeds testing various mitigation strategies are resolved nowadays with ABMs. In particular, recently ABMs have been used extensively to inform public health in (non-pharmaceutical) interventions against the spread of COVID-19 ~\cite{2020Ferguson,eubank_commentary_2020,lanl2020covid,2020ABM-calibration,kaxiras2020multiple}, and verify new strategies like test-trace-quarantine \cite{ABM-Gates}, among many other applications. 

There are two major problems with the modeling of pandemic.  First, many parameters need to be calibrated on data.  Second, even when calibrated for the current state of pandemic the models which are too detailed become impractical for making a forecast and for developing prevention strategies -- both requiring checking multiple (forecast and/or prevention) scenarios. Using ABMs, which are clearly over-modeled (too detailed) is especially problematic in the context of the latter. 
For example, the open-source ABM solver FLUTE~\cite{chao2010flute} developed originally for modeling influenza, works with data that are acquired through Geographic Information Systems (GIS) on the scale of census tracts or communities, which is a very reasonable scale of spatial resolution to understand the dynamics of pandemics on a local scale. FLUTE populates each of the communities with thousands to millions of inhabitants in order to account for their daily patterns of travel. We believe that constructing effective Graphical Models (GM) of Pandemics with community-scale spatial resolution and then modeling pairwise (and possibly higher-order) epidemic interactions between communities directly, without introducing the thousands-to-millions of dummy agents, will complement (as discussed in the next paragraph), but also improve upon ABMs by being more efficient, robust and easier to calibrate.

An important, and possibly one of the first, Graphical Model (GM) of the COVID-19 pandemic was proposed in \cite{chang_mobility_2020}. Dynamic bi-partite GMs connecting census tracts to specific Points Of Interest (non-residential locations that people visit such as restaurants, grocery stores and religious establishments) within the city and studying dynamics of the four-state (Susceptible, Exposed, Infectious and Removed) of a census tract (graph node) on the graph, were constructed in \cite{chang_mobility_2020}  for major metro-area in USA based on the SafeGraph mobility data \cite{SafeGraph-Mobility}.

\begin{figure}[ht]
\centering
\includegraphics[width=0.2\textwidth]{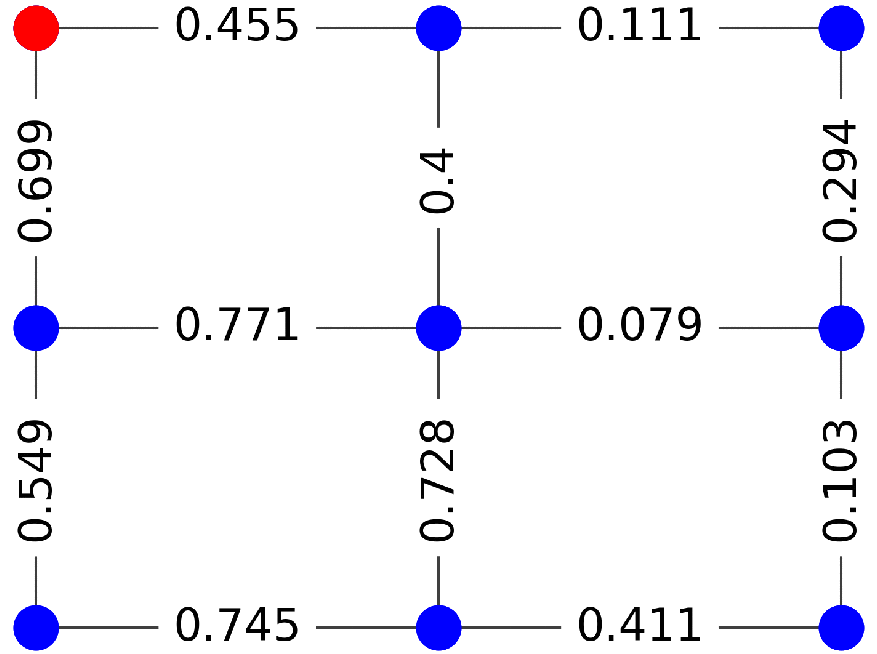}\hfill
\includegraphics[width=0.2\textwidth]{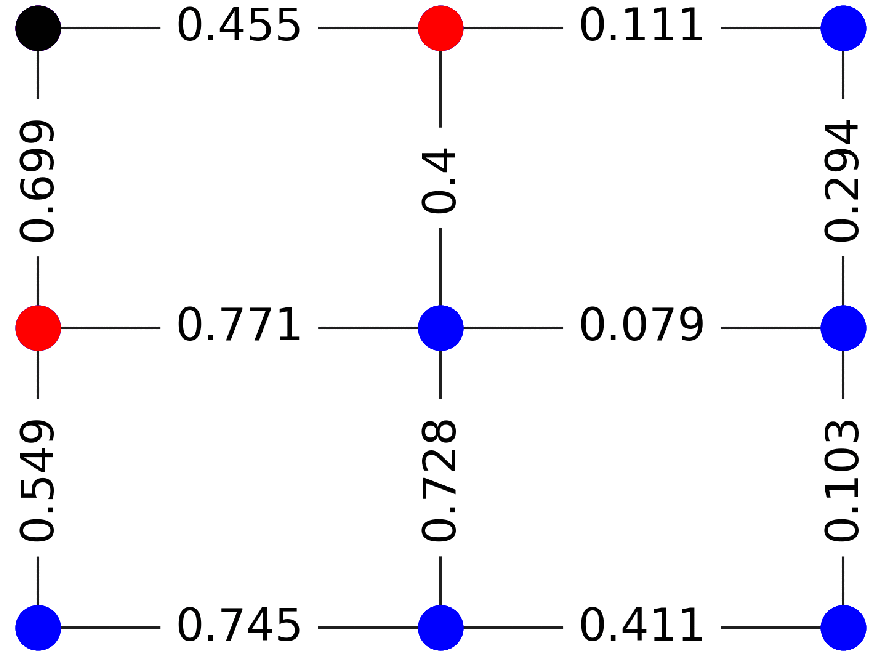}\hfill
\includegraphics[width=0.2\textwidth]{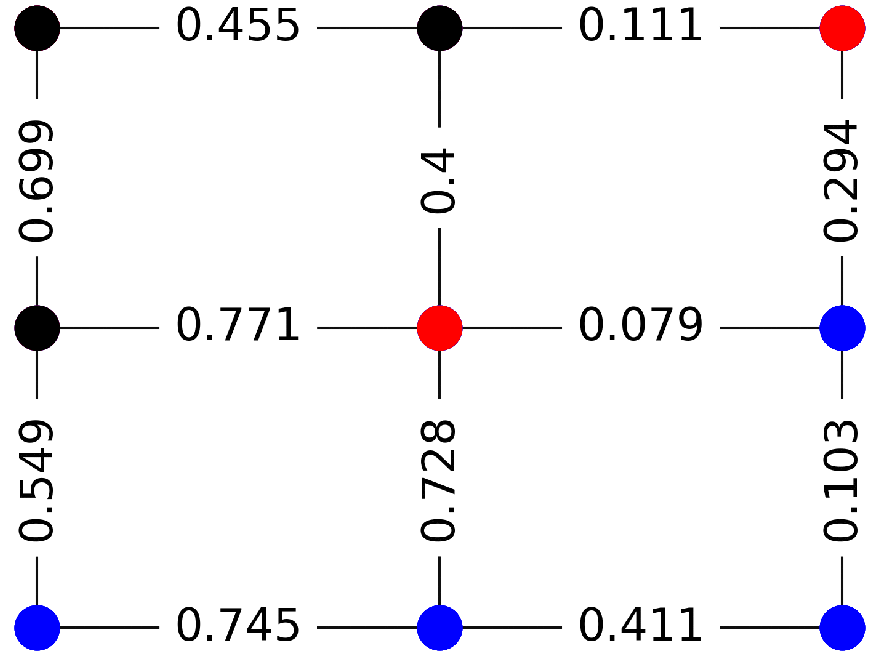}\hfill
\includegraphics[width=0.2\textwidth]{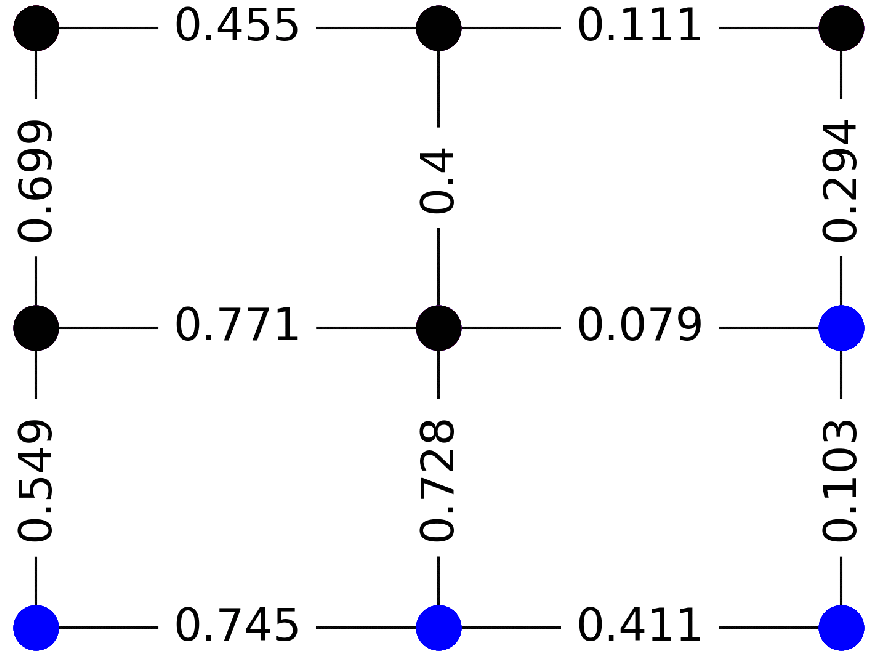}
% Reduce the figure size so that it is slightly narrower than the column.
\caption{An exemplary random sequence (top-left to top-right to bottom-left to bottom-right) of the Independent Cascade Model (ICM) dynamics over $3\times3$ grid. Nodes colored red, blue, and black are {\bf I}nfected, {\bf S}usceptible, and {\bf R}emoved at the respective stage of the dynamical process. This (shown) sample of the dynamic process terminates in 3 steps. Ising Model of Pandemic (IMP), which is the focal point of this manuscript, describes a regularized version of the ICM terminal state, where only two states ({\bf S}-blue and {\bf R}-black) are left. (See text for details.)}
\label{fig:ICM}
\end{figure}

In fact, similar dynamic GMs, e.g. of the Independent Cascade Model (ICM) type  \cite{2003KempeKleinbergTardos,2012NetrapalliSanghavi,2012GomezLeskovecKrause,KhaDilSon13,2016RosenfeldNitzanGloberson}, % {\color{red}Place a citation here\cite{}},  
were introduced even earlier in the CS/AI literature in the context of modeling how the rumors spread over social networks (with a side reference on using ICM in epidemiology). As argued in \cite{GM-Pandemic} the Independent Cascade Models (ICMs) can be adapted to modeling pandemics. (Another interesting use of the ICM to model COVID-19 pandemic was discussed in \cite{Chen_2020}.) In its minimal version, an ICM of Pandemic can be built as follows. Assume that the virus spreads in the community (census tract) sufficiently fast, say within five days -- which is the estimate for the early versions of COVID-19 median incubation period. If an infected person enters a community/neighborhood but does not stay there,  he infects others with some probability. If a single resident of the community becomes infected, all other residents are assumed infected as well (instantaneously). The model is a discrete-time dynamic model in which nodes in a network are in one of the three states: {\bf S}usceptible, {\bf I}nfected, or {\bf R}emoved. The nodes represent communities/neighborhoods. 
A contact between an {\bf I}nfected community/node and another community which is {\bf S}usceptible has an assigned probability of disease transmission, which can also be interpreted as the probability of turning the {\bf S} state into {\bf I} state.  Consistently with what was described above, the network is represented as a graph, where nodes are tracts and edges, connecting two tracts, have an associated strength of interaction representing the probability for the infection to spread from one node to its neighbor. A seed of the infection is injected initially at random, for example, mimicking an exogenous super-spreader infection event in the area; examples could include political or religious gatherings. See Figure~\ref{fig:ICM} illustrating dynamics of the cascade model over 3-by-3 grid  graph. Color coding of nodes is according to {\bf S}usceptible=blue, {\bf I}nfected=red, {\bf R}emoved=black. Given the starting infection configuration, each infected community can infect its graph-neighbor community during the next time step with the probability associated with the edge connecting the two communities. Then the infected community moves into the removed state. The attempt to infect each neighbor is independent of all other neighbors. This creates a cascading spread of the virus across the network. The cascade stops in a finite number of steps, thereby generating a random {\bf R}emoved pattern, shown in black in the Fig.~\ref{fig:ICM},  while other communities which were never infected (remain {\bf S}usceptible) are shown in blue. 

It was shown in \cite{GM-Pandemic} that with some regularization applied, statistics of the terminal state of the Cascade Model of Pandemic turns into a Graphical Model of the attractive Ising Model type. 

\underline{\bf This manuscript Road Map.}  Working with the Ising Model of Pandemic, we start the technical part of the manuscript by posing the {\bf Inference/Prediction Challenge} in Section {\bf Ising Model of Pandemic}. Here,  the problem is stated,  first, as the Maximum A-Posteriori over an attractive Ising Model,  and we argue,  following the approach which is classic in the GM literature,  that problem can be re-stated as a tractable LP. We then proceed to Section {\bf Prevention Challenge} to pose the main challenge addressed in the manuscript -- the {\bf Prevention Challenge} -- as the two-level optimization with inner step requiring resolution of the aforementioned Prediction Challenge. Aiming to reduce the complexity of the Prevention problem, we turn in Section {\bf Geometry of the MAP States} to the analysis of the conditions in the formulation of the Prediction Challenge, describing the Safety domain in the space of the Ising Model parameters. We show the Safety domain is actually a polytope, even though exponential in the size of the system.  We proceed in Section {\bf Projecting to the Safe Polytope} with analysis of the Prevention Challenge,  discussing the interpretation of the problem as a projection to the Safety Polytope from the polytope exterior, needed when the bare prediction suggests that system will be found with high probability outside of the Safety Polytope. Section {\bf Two Polarized Modes} is devoted to approximation which allows an enormous reduction in the problem complexity. We suggest here that if the graph of the system is sufficiently dense, the resulting MAP solution may only be in one of the two polarized states (a) completely safe (no other nodes except the initially infected) pick the infection, or (b) the infection is spread over the entire system. We support this remarkable simplification by detailed empirical analysis and also by some theoretical arguments. Section {\bf Results} is devoted to the experimental illustration of the methodology on the practical example of the Graphical Model of Seattle. The manuscript is concluded in Section {\bf Discussion} with a brief summary and discussion of the path forward.

{  We conclude this introductory Section with a general comment about relation between ABM and GM approaches. %{  (We are thankful to anonymous review for raising the important point.)}  
Focusing on GMs we did not mean to suggest that the GM approach is outright better than the ABM approach.  In fact, we do believe that the two are complementary.  ABM is more accurate but computationally heavy, while GM is coarser but computationally lighter.  We are also convinced that it will be advantageous in the future to consider the two in tandem,  e.g.,  with the parameters of the GM trained on the synthetic data generated by the ABM.}

\section*{Ising Model of Pandemic}
\label{ref:IMP}

As argued in \cite{GM-Pandemic} the terminal state of a dynamic model generalizing the ICM model can be represented by the Ising Model  of Pandemic (IMP), defined over graph $\cal G = ({\cal V}, {\cal E})$, where ${\cal V}$ is the set of $N=|{\cal V}|$ nodes and ${\cal E}$ is the set of undirected edges. The IMP, parameterized by the vector of the node-local biases, $h=(h_a|a\in\cal V)\in \mathbb{R}^N$, and by the vector of the pair-wise (edge) interactions, $J=(J_{ab}|\{a,b\}\in {\cal E})$, 
describes the following Gibbs-like probability distribution for a state, $x=(x_a=\pm 1|a\in{\cal V})\in 2^{|\cal V|}$, associated with $\cal V$:
\begin{equation}
P(x~ \vert~ J, h) = \frac{\exp{(-E(x~ \vert~ J, h))}}{Z(J, h)}, \label{eq:Gibbs}
\end{equation}
where any node, $a\in \cal V$ can be found in either {\bf S}- (susceptable, never infected) state,  marked as $x_a=-1$, or {\bf R}- (removed, i.e. infected prior to the termination) state, marked as $x_a=+1$.  In Eq.~(\ref{eq:Gibbs}), $E(x| J, h)$ and $Z(J, h)$ are model's energy function and partition function respectively:
\begin{eqnarray}\label{eq:E}
E(x~ \vert~ J, h) &=& \sum\limits_{a\in \cal V} h_a x_a - \sum\limits_{a, b\in \cal V} J_{ab} x_a x_b,\\ \label{eq:pf}
Z(J, h) &=& \sum\limits_{x} \left (\sum\limits_{a\in \cal V} h_a x_a - \sum\limits_{a, b\in \cal V} J_{ab} x_a x_b \right ). 
\end{eqnarray}

In what follows, we will focus on finding the Maximum-A-Posteriori (MAP) state of the IMP conditioned to a particular initialization -- setting a subset of nodes, ${\cal I}\in {\cal V}$,  to be infected. We coin the MAP problem {\bf Inference Challenge}:

\section*{Inference Challenge}

In this paper, we focus on finding the mode of distribution defined in eq.~\ref{eq:Gibbs}. This problem is called maximum a posteriori (MAP) estimation, and the most probable states are called MAP states, or ground states (in physics literature). Moreover, in our setting, some nodes are initialized to plus one state (we denote this set of nodes by $I$). The corresponding binary optimization problem is formulated as follows:

\begin{gather}\label{eq:InfQ}
x^{(\text{MAP})}({\cal I}~ \vert~ J,h)=\text{arg}\min\limits_{x} E(x~ \vert~ J, h)\Big|_{\forall a\in {\cal I}:\quad  x_a = +1},
\end{gather}
where we emphasize dependence of the MAP solution on the set of the initially infected nodes, ${\cal I}$. 

Note that in general finding $x^{(\text{MAP})}$ is NP-hard \cite{Barahona_1982}.  However  if $J>0$ element-wise, i.e. the Ising Model is attractive (also called ferromagnetic in statistical physics), Eq.~(\ref{eq:InfQ}) becomes equivalent to a tractable (polynomial in $N$) Linear Programming (see \cite{14ZWP} and references therein).  In fact, the IMP is attractive, reflecting the fact that the state of a node is likely to be aligned with the state of its neighbor. 

Let us also emphasize some other features of the IMP:
\begin{enumerate}
    \item $\cal G$ should be thought of as an "interaction" graph of a city,  reflecting transportation, commutes, and other forms of interactions between populations with the homes at the two nodes (census tracts) linked by an edge. The strength of a particular $J_{ab}$ shows the level of interaction associated with the edge $\{a,b\}$. 
    \item A component, $h_a$, of the vector of local biases, $h$, is reflecting $a$-node specific factors such as immunization level, imposed quarantine, and degree of compliance with the public health measures (e.g., wearing masks and following other rules). Large negative/positive $h_a$ shows that residents of the census tract associated with the node $a$ are largely healthy/infected. 
\end{enumerate}
If solution of the Inference Challenge problem is such that the ${\bf R}$-subset of the MAP solution, $x^{(\text{MAP})}({\cal I}|J,h)$, i.e. 
\begin{gather}\label{eq:V-I}
    {\cal R}({\cal I},J,h)=\Big\{a\in {\cal V} \, | \, x_a^{(\text{MAP})}({\cal I}|J,h)=+1\Big\},
\end{gather}
is sufficiently large, we would like to mitigate the infection, therefore setting the Prevention Challenge discussed in the next Section.

\section*{Prevention Challenge}\label{sec:Prevention}

Let us assume that modification of $J$ and $h$ are possible and consider the space of all feasible $J$ and $h$. We will then identify {\it Safe Domain} as a sub-space of feasible $J$ and $h$ such that for all the initial sets of the initially infected nodes, ${\cal I}$, considered the resulting "infected" subset, ${\cal R}({\cal I},J,h)$, is sufficiently small. A more accurate definition of the Safe Domain follows.  Then, we rely on the definition to formulate the control/mitigation problem coined Prevention Challenge. At this stage, we would also like to emphasize that studying the geometry of the Safe Domain is one of the key contributions of this manuscript.

{\bf Definition.} 
Consider IMP over ${\cal G}=({\cal V},{\cal E})$ and  with the parameters $(J,h)$.  Let us also assume that the set of initially infected nodes, ${\cal I}$, is drawn from the list, $\Upsilon$. We say that $(J, h)$ is in the $k$-{\it Safe Domain} if for every ${\cal I}$ from $\Upsilon$ the number of ${\bf R}$-nodes in the MAP solution (\ref{eq:InfQ}), is at most $k$, i.e. 
\begin{gather}\label{eq:k-safe}
    \forall {\cal I}\in \Upsilon:\quad |{\cal R}({\cal I},J,h)|\leq k,
\end{gather}
where ${\cal R}({\cal I},J,h)$ is defined in Eq.~(\ref{eq:V-I}). 

{\bf Prevention Challenge:} Given $(J^{(0)},h^{(0)})$ describing the bare status of the system (city) which is not in the $k$-{\it Safe Domain}, and given the cost of the $(J,h)$ change, $C\left((J,h);(J^{(0)},h^{(0)})\right)$, what is least expensive change to $(J^{(0)},h^{(0)})$ state of the system which is in the the $k$-{\it Safe Domain}? Formally,  we are interested to solve the following optimization:
\begin{gather}\label{eq:prevention-Q}
    (J^{(\text{\tiny corr})},h^{(\text{\tiny corr})})=\text{arg}\min\limits_{(J,h)} C\left((J,h);(J^{(0)},h^{(0)})\right)_{\text{Eq.~(\ref{eq:k-safe})}}.
\end{gather}

Expressing it informally, the Prevention Challenge seeks to identify a minimal correction (thus "corr" as the upper index) $(J^{(\text{\tiny corr})},h^{(\text{\tiny corr})})$, which will move the system to the safe regime from the unsafe bare one, $(J^{(0)},h^{(0)})$.  The measures may include limiting interaction along some edges of the graph, thus modifying some components of $J$, or enforcing local biases, e.g., increasing level of vaccination, at some component of $h$.

Given that condition in Eq.~(\ref{eq:k-safe}) also requires solving Eq.~(\ref{eq:InfQ}) for each candidate $(J,h)$, the {\bf Prevention Challenge} formulation is a difficult two-level optimization.
However, as we will see in the next Section, the condition in Eq.~(\ref{eq:k-safe}) (and thus the inner part of the aforementioned two-level optimization) can be re-stated as the requirement of being inside of a polytope in the $(J,h)$ space. In other words, the $(k)$-{\it Safe Domain} is actually a polytope in the $(J,h)$ space. 

\section*{Geometry of the MAP States}\label{sec:Geometry}

Before solving the Prevention Challenge problem, we want to shed some light on the geometry of the MAP states. We work here in the space of all the Ising models over a graph ${\cal G}=({\cal V},{\cal E})$, where each of the models is specified by $(J,h)$. 

{\bf Proposition.} Safe Domain of a graph ${\cal G}=({\cal V},{\cal E})$ with $N=|{\cal V}|$ nodes is a polytope in the space of all feasible parameters, $(J,h)$, defined by an exponential in $N$ number of linear constraints.

{\bf Remark.} The Proposition allows us, from now on, to use {\it Safe Polytope} instead of the {\it Safe Domain}.

{\bf Proof of the Proposition.} The space of all the Ising models is divided into $2^N$ regions by the corresponding MAP states. Moreover, the boundary between any pair of neighboring regions is linear: consider two states $x^{(i)}$ and $ x^{(j)}$, and denote $(J,h)^{(i)}$ (resp.  $(J,h)^{(j)}$) the set of all the Ising models with the MAP state $x^{(i)}$ (resp. $x^{(j)}$), then $(J,h)^{(i)}$ and $(J,h)^{(j)}$ are separated by the equation, $E(x^{(i)}~ \vert~ J,h) = E(x^{(j)}~ \vert~ J,h)$, which is linear in $(J,h)$. For a subset, $R\subseteq \cal V$, of nodes, let $x^{(R)}$ be the state in which, $x_a=+1,\ \forall a\in R, x_a=-1,\ \forall a\notin R$. Let $X^{(R)}$ be the set of all the MAP states, $x$, such that $\forall a\in R,\ x_a=+1$ (while other nodes, i.e. $b\in {\cal V}\setminus R$, are not constrained, $x_b=\pm 1$). Then the $k$-Safe Polytope, which we denote, $\text{SP}(k)$, is defined by at most $\sum\limits_{k^\prime=1}^{k}\binom{N}{k^\prime}\cdot(2^{N-k^\prime}-1) $ linear inequalities:
\begin{equation}\label{eq:SP-k}
\text{SP}(k) = \hspace{-0.9cm}\bigcap\limits_{\begin{array}{c}\forall R,|R|\le k;\\ \forall x\in X^{(R)}\setminus x^{(R)}\end{array}}\hspace{-1.05cm} \left\{(J, h)~ \vert~ E(x^{(R)} | J, h) > E(x~ \vert~ J, h)\right\},
\end{equation}
were some of these linear inequalities on the right  hand side may be redundant.

%Alon changed rand hand ->right hand

{\bf Remark.} In the case of $k=1$ (which, obviously, applies only if all the initial infections are at a single nodes, i.e. $\forall {\cal I}\in \Upsilon,\ |{\cal I}|=1$), there are at most, $N \cdot (2^{N-1} - 1)$ linear inequalities.

\begin{figure}[ht]
\centering
\begin{subfigure}{.45\textwidth}
    \centering
    \includegraphics[width=0.9\columnwidth]{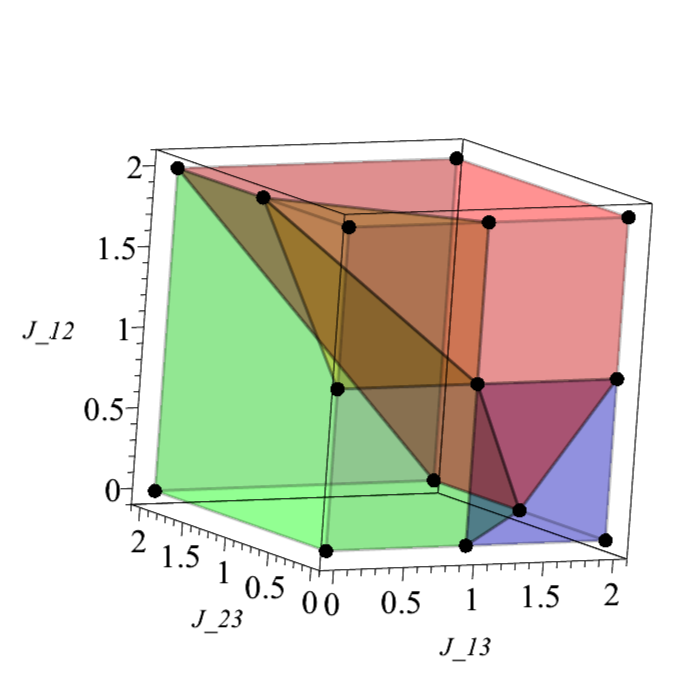} 
    \caption{Geometry of the attractive Ising model illustrated  on the example of a triangle graph ($K_3$) when a single node is infected. See explanations in the text.}
    \label{fig:geometry}
    \end{subfigure}%
\hfill
\begin{subfigure}{.45\textwidth}
    \centering
    \includegraphics[width=0.9\columnwidth]{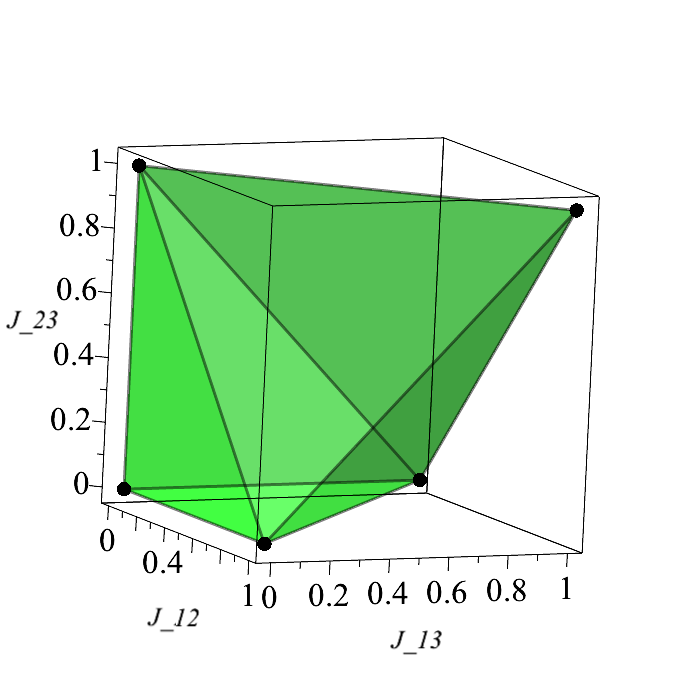} 
    \caption{The Safe Polytope illustrated on the example of a triangle graph ($K_3$) with field vector $h = [-1, -1, -1]$. See explanations in the text.}
    \label{fig:SafePoly}
\end{subfigure}
\caption{Geometry of MAP states for Ising on $K_3$.}
\label{fig:SafePoly_both}
\end{figure}

We illustrate the geometry of the Ising model over the triangle graph (three nodes connected in a loop, $K_3$) in Fig.~\ref{fig:geometry} and Fig.~\ref{fig:SafePoly}. For both illustrations, we fix the $h$ value to $-1$ at all the nodes, and we are thus exploring the remaining three degrees of freedom, $J_{12},J_{13},J_{23}$ (since $J$ is symmetric),
which corresponds to exploring interactions within the class of attractive Ising models, $\forall a,b=1,2,3:\ J_{ab}\in \mathbb{R}_+$. 

First, we consider the case when the only node $a=1$ is infected. In this simple setting there are four possible MAP states, $(x_1,x_2,x_3)\in \{(+1, -1, -1), (+1, -1, +1), (+1, +1, -1),(+1,+1,+1)\}$, shown in Fig.~(\ref{fig:geometry}) as green, blue, yellow and red, respectively. Finally, in the figure Fig.~(\ref{fig:SafePoly}) we plot the Safe Polytope $\text{SP}(1)$.
We observe that the two "polarized" MAP states, $(+1,-1,-1)$ and $(+1,+1,+1)$, are seen most often among the samples, while domain occupied by the other two "mixed" MAP states, $(+1,-1,+1)$ and $(+1,+1,-1)$ is much smaller, with the two modes  positioned on the interface between the two polarized states.

As will be shown below in the next Section, the polarization phenomena with only two "polarized" MAP states, which we coin in the following the two polarized modes, which we see on this simple triangle example, is generic for the attractive Ising model.

\section*{Two Polarized Modes}\label{sec:Polarized}

{\bf Definition.} Consider a particular subset of the initially infected nodes, ${\cal I}$ (where thus, $\forall a\in{\cal I}:\ x_a=+1$). We call the MAP state of the model  {\it polarized} when one of the following is true:  {\bf (i) }  only initially infected nodes show $+1$ within the MAP solution, $\forall a\in {\cal I}:\ x_a=+1,\ \forall b\in{\cal V}\setminus{\cal I}:\ x_b=-1$ or {\bf (ii)} all nodes within the MAP state show $+1$, $\forall a\in{\cal V}:\ x_a=+1$. We call a MAP state {\it mixed} otherwise.

Experimenting with many dense graphs, which are typical in the pandemic modeling of modern cities with extended infrastructures and multiple destinations visited by many inhabitants, we observe that the two polarized MAP states dominate generically,  while the mixed states are extremely rare. 

\begin{figure}[ht]
\centering
\begin{subfigure}{.45\textwidth}
    \centering
    \includegraphics[width=0.9\columnwidth]{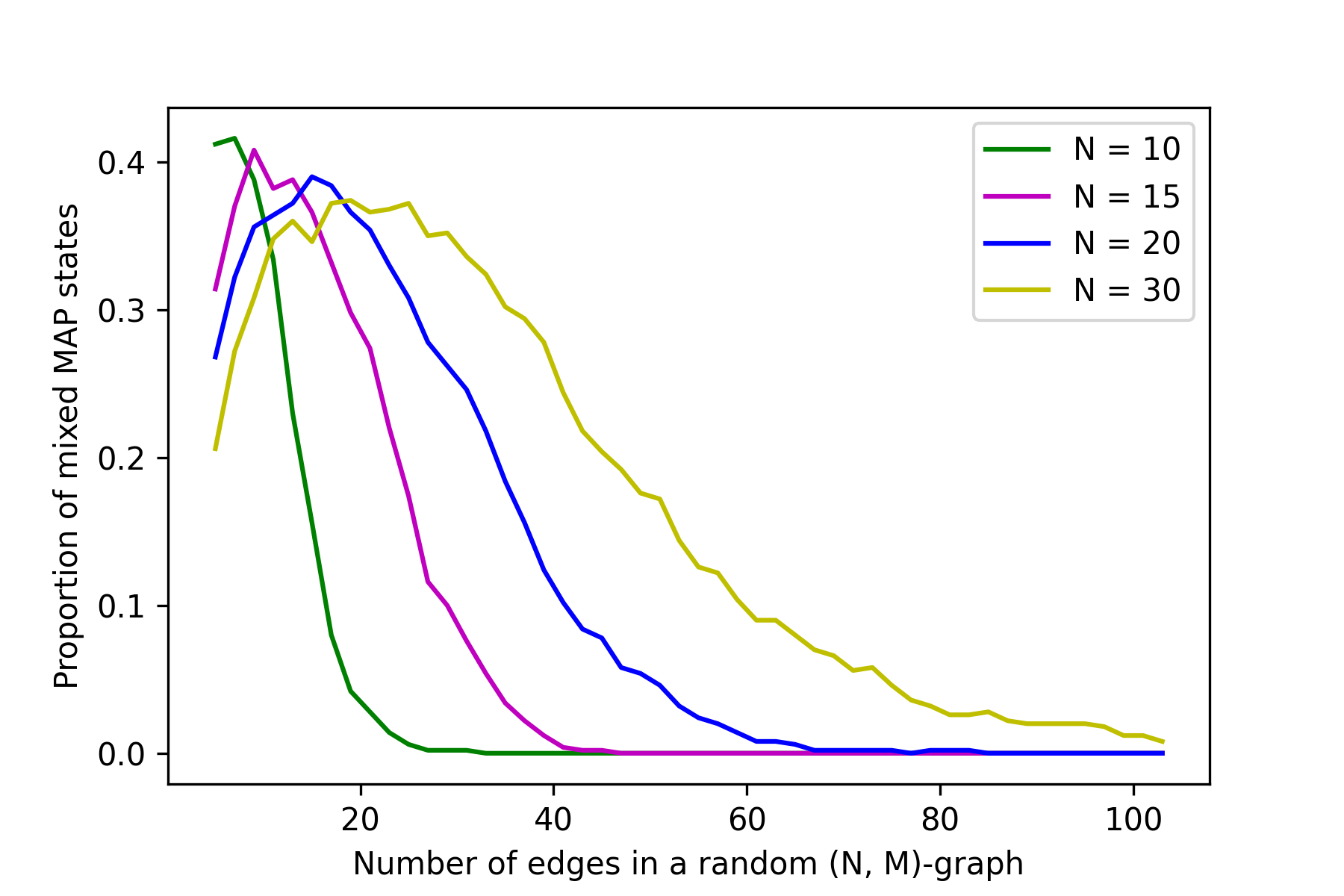} 
    \caption{Proportion of the mixed states in all samples for an ensemble of the (attractive) Ising Model of Pandemic over graphs with $N$ nodes, shown as a function of the varying number of edges, $M$. Each shown point is the result of the averaging over $500$ random instances of the $(J,h)$ over the same graph. (See text for additional details.)} 
    \label{fig:density}
\end{subfigure}%
\hfill
\begin{subfigure}{.45\textwidth}
    \centering
    \includegraphics[width=0.9\columnwidth]{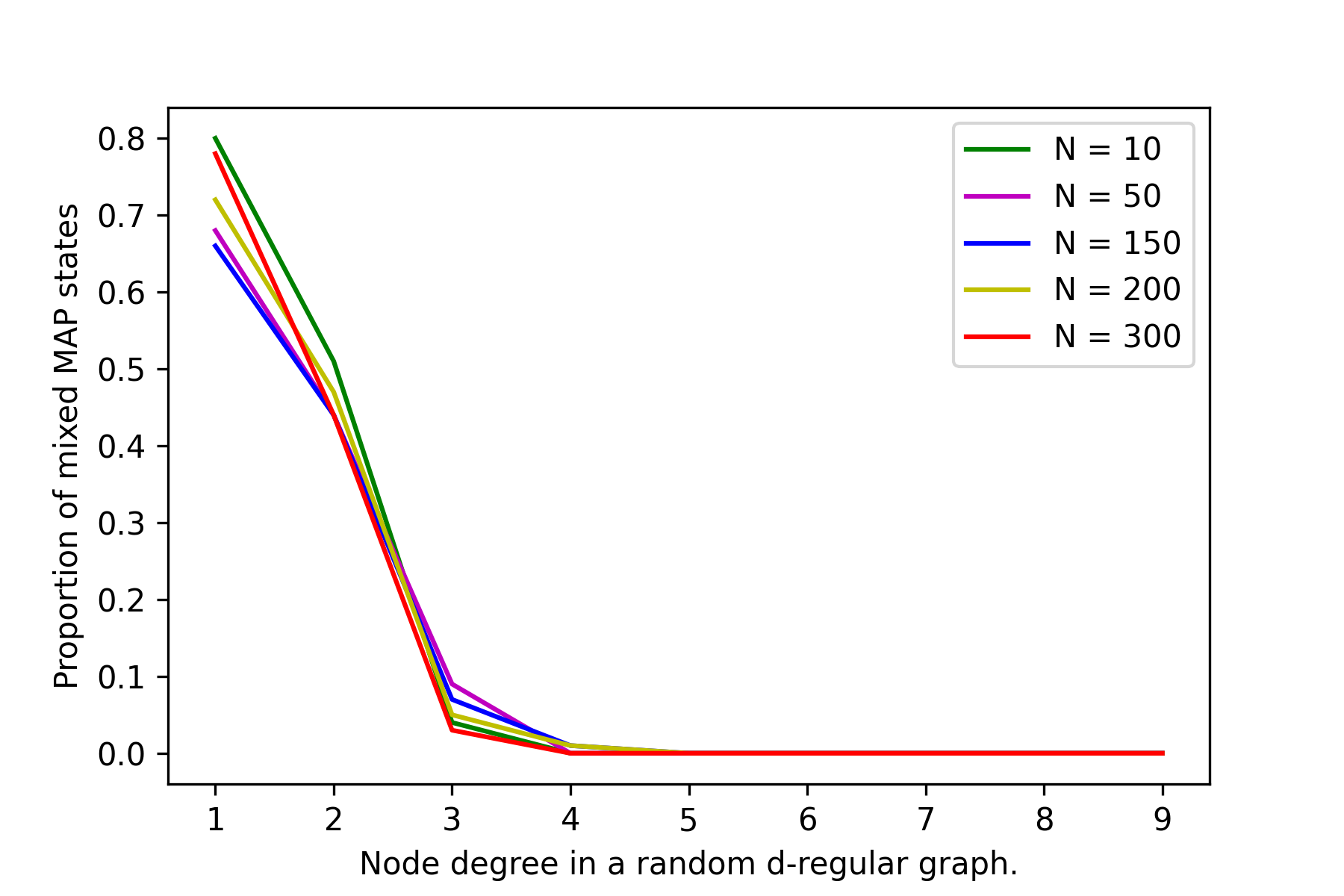} 
    \caption{Proportion of the mixed-to-polarized states {  ($y$-axis) vs $\alpha$ ($x$-axis)} for an ensemble of the (attractive) Ising Model of Pandemic over $d$-regular graphs with $N$ nodes, shown as a function of $d$. Each point is the result of  averaging over $100$ random instances of the $(J,h)$ over different random graphs with the same node degree. (See text for additional details.)
    }
    \label{fig:density_p}
\end{subfigure}
\caption{Mixed-to-Polarised Transition Experiments.}
\label{fig:density_both}
\end{figure}

Fig.~\ref{fig:density} illustrates results of one our ensemble of random IMPs' experiments. We, first, fix $N$ to $20$, pick $M$ such that $M\le N(N-1)/2=190$ and then generate at random $M$ edges connecting the $20$ nodes. 
Then, for each of the random graphs (characterized by its own $M$) we generate $500$ random samples of $(J,h)$, representing attractive Ising models. Finally, we find the MAP state for each IMP instance, count the number of mixed states and show the dependence of the fractions of the mixed states (in the sample set) in the Fig.~(\ref{fig:density}). A fast decrease of the proportion of the mixed states is observed with an increase in $M$.

Extension of these experiments (see Fig.~(\ref{fig:density_p})) suggests that when we consider an ensemble of IMPs over graphs with $N$ nodes and the average degree $\alpha=O(1)$ which is sufficiently large (so that the graph is sufficiently dense) and increase $N$, we observe that the Mixed State Probability (MSP), or equivalently proportion of the mixed-to-polarized states, decreases dramatically. Moreover, based on the experiments, we conjecture that the MSP decays to zero at $\alpha>\alpha_c$, but it saturates at $\alpha<\alpha_c$,  where $\alpha_c$ is the threshold depending on the ensemble details.  This threshold behavior is akin to the phase transition that occurred in many models of the spin glass theory \cite{MezardParisiVirasoro} and many models of the Computer Science and Theoretical Engineering defined over random graphs and considered in the thermodynamic limit, i.e. at $N\to \infty$. See e.g. \cite{RichardsonUrbanke} (application in the Information Theory, and specifically in the theory of the Low Density Parity Check Codes) \cite{MezardMontanari} (applications in the Computer Science, and specifically for random SAT and related models) and references therein. We postpone further discussions of the conjecture for a future publication (see also brief discussion in Section {\bf Discussion}). 

We will continue discussion of the two-mode solution in the next Section. 

\section*{Projecting to the Safe Polytope}\label{sec:Projection}

In this Section we aim to summarize all the findings so far to resolve the Prevention Challenge formulated in Section {\bf Prevention Challenge}, specifically in  Eq.~(\ref{eq:prevention-Q}) stating the problem as finding a minimal projection to the Safety Domain/Polytope from its exterior. The task is well defined, but in general, and as shown in Section {\bf Geometry of the MAP States}, it is too complex  -- as the description of the Safety Polytope (number of linear constraints, required to define it) is exponential in the system size (number of nodes in the graph).  However, the two-mode approximation, introduced in Section {\bf Two Polarized Modes}, suggests a path forward: use the two-mode approximation and therefore remove all the linear constraints but one, separating the two polarized states. 

Let us denote the two-mode approximation of the Safe Polytope by $\widehat{SP}(\Upsilon)$, where thus $k$ in the original Safe Polytope, $SP(k)$, is replaced by the set $\Upsilon$ of all the initial infection patters. Then we write,
\begin{gather}\label{eq:SP-Upsilon}
    \widehat{SP}(\Upsilon)=\bigcap\limits_{{\cal I}\in \Upsilon}\big\{(J, h)~ \vert~ E(+1^{2N} ~ \vert~ J, h) \geq  E(x^{({\cal I})}~ \vert~ J, h)\big\},
\end{gather}
where, $\forall a\in{\cal I}:\ x^{({\cal I})}_a=+1$ and $\forall b\in {\cal V}\setminus {\cal I}:\ x^{({\cal I})}_b=-1$. Eq.~(\ref{eq:SP-Upsilon}) represents a polytope stated in terms of the $|\Upsilon|$ constraints. In particular, if $\Upsilon$ accounts for all the initial infections, ${\cal I}$, of size not large than $k$, then $|\Upsilon|=\sum_{k^\prime=1}^k \binom{N}{k^\prime}$: the number of constraints grows exponentially in the maximal size of the initial infections, however the number of the constraints remains tractable for any $k=O(1)$. Replacing conditions in Eq.~(\ref{eq:prevention-Q}) by  $\widehat{SP}(\Upsilon)$, defined in  Eq.~(\ref{eq:SP-Upsilon}), one arrives at the following tractable (in the case of $k=O(1)$) convex optimization expression answering the Prevention Challenge approximately (within the two-mode approximation):
\begin{gather}\label{eq:prevention-Q-two-mode}
    (\widehat{J}^{(\text{\tiny corr})},\widehat{h}^{(\text{\tiny corr})})=\text{arg}\min\limits_{(J,h)} C\left((J,h);(J^{(0)},h^{(0)})\right)_{\text{Eq.~(\ref{eq:SP-Upsilon})}}.
\end{gather}
This formula is the final result of this manuscript analytic evaluation. 
In the next Section we use Eq.~(\ref{eq:prevention-Q-two-mode}), with $C(\cdot;\cdot)$ substituted by the {  $l_1$-norm, $l_2$-norm and their convex combination,} to present the result of our experiments in a quasi-realistic setting describing a (hypothetical) pandemic attack and optimal defense, i.e., prevention scheme.

\section*{Results}\label{sec:Experiments}

\subsection*{Seattle data}

\begin{figure}[t]
\centering
\includegraphics[width=0.65\columnwidth]{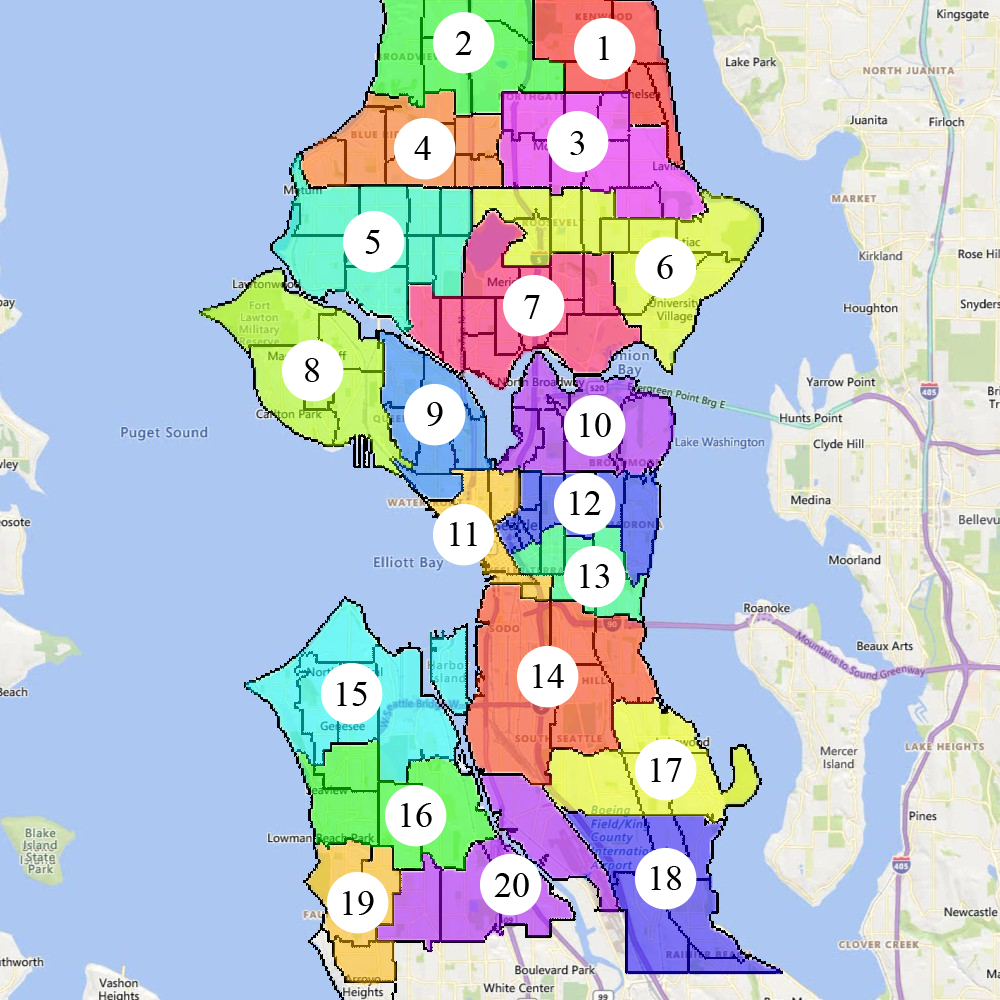}
\caption{Seattle case study areas and census tracts \cite{seattleCensusTractMap2010}.}
\label{fig:seattle20area}
\end{figure}

We illustrate our methodology on a case study of the city of Seattle. {  (To verify scalability of the approach we have also experimented with larger system, e.g., models of New-York City and of the State of Wisconsin. This is still work in progress, partially reported in \cite{supplement}.)} Seattle has 131 Census Tracts. Each Census Tract includes 1 to 10 Census Block Groups with  600 to 3000 residents and represents 1200 to 8000 population.  Boundaries separating Census Tracts are designed to represent natural or urban landmarks and also to be persistent over time \cite{USBureauGlossary}. To reduce complexity, we merge census tracts into 20 regions. See Fig.~~\ref{fig:seattle20area}. To prepare this splitting of Seattle into 20 regions/nodes, we utilize geo-spatial information from the TIGER/Line Shapefiles project provided by U.S. Census Bureau \cite{USBureauTiger}. The travel data of Seattle was extracted from the Safegraph dataset \cite{SafeGraph-DataConsortium}, which provides anonymized mobile tracking data. Each data point in the Safegraph database describes the number of visits from a Census Block to a specific point of interest represented by latitude and longitude. Mobility data associated with travelers crossing the boundaries of Seattle was ignored. 

We then follow the methodology developed in \cite{GM-Pandemic} to combine the aggregated travel data with the epidemiological data, representing current state of infection in the area. This results in the estimation of the pair-wise interactions, $J$, parameterizing the Ising Model of Pandemic. We also come up with an exemplary (uniform over the system) local biases, $h$, completing the definition of the model.  (We remind that the prime focus of the manuscript is on developing methodology which is AI sound and sufficiently general. Therefore,  the data used in the manuscript are roughly representative of the situation of interest, however not fully practical.) We consider a situation with different levels of infection and chose $(J^{(0)},h^{(0)})$ stressed enough,  that is resulting in the prediction (answer to our Prediction Challenge), which lands the system in the dangerous domain -- outside of the Safety Polytope. 

\subsection*{Convex projection}

In all of our experiments, we have used the general-purpose Gurobi optimization solver \cite{gurobi} to compute the MAP states and thus to validate the two-mode assumption. (We have also experimented with CVXPY \cite{cvxpy}, but found it performing slower than Gurobi, at least over the relatively small samples considered in this proof of principles study. In the future, we plan to use existing, or developing new, LP solvers designed specifically for finding the MAP state of the attractive Ising model.) To illustrate our Prevention strategy, we took the Seattle data described above, and fed it as an input into the optimization (\ref{eq:prevention-Q-two-mode}), describing projection to the Safety Polytope, where $C(\cdot;\cdot)$ is substituted by the $l_1$ norm. CVXPY solver was used for this convex optimization task. Our code (python within jupyter notebook) is available at \cite{supplement}.

Table \ref{tbl:projection} shows  results of our Prevention experiments on the Seattle data. We analyze{ , first,} $l_1$ projection to $\widehat{SP}(\Upsilon)$ where $\Upsilon$ consists of all the initial infection patterns consisting of up to $k$ nodes. In all of our experiments, the values of the field vector $h$ (uniform across the system) was fixed to $-1$. We observe that the number of constraints grows exponentially with $k$; however, the cost of intervention remains roughly the same. We intend to analyze the results of this and other (more realistic) experiments in future publications aimed at epidemiology experts and public health officials. 

\begin{table}[ht]

\centering
\begin{tabular}{l|l|l|l}
%\hline
k & LP Constraints & Runtime & Cost\\
\hline
1 & 801 &  1.65s   &   41.69  \\
\hline
2 & 991 &  3.04s   &  43.62  \\
\hline
3 &  2131 &  10.90s   &   44.30  \\
\hline
4 &  6976 &  100.08s  &   44.56  \\
\hline
%\bottomrule
\end{tabular}
\caption{Summary of our prevention experiments on the Seattle data. $k$, in the first column, is the maximal number of nodes in the initially infected patterns (all accounted for to construct the $k$-Safe Polytope). The second column shows number of linear constraints characterizing the $k$-Safe Polytope. Respective Run Time and Cost are shown in the 3rd and 4th column,  where Cost shows the difference in $l_1$ norm between the $(J^{(0)},h^{(0)})$, characterizing stressed but unmitigated regime, and the optimal prevention regime, resulting in $(\widehat{J}^{(\text{\tiny corr})},\widehat{h}^{(\text{\tiny corr})})$ computed according to Eq.~(\ref{eq:prevention-Q-two-mode}).}
\label{tbl:projection}
\end{table}

\begin{figure}[ht]
\centering
    \begin{subfigure}[b]{0.49\textwidth}
         \centering
         \includegraphics[height=0.34\textheight,width=0.99\columnwidth]{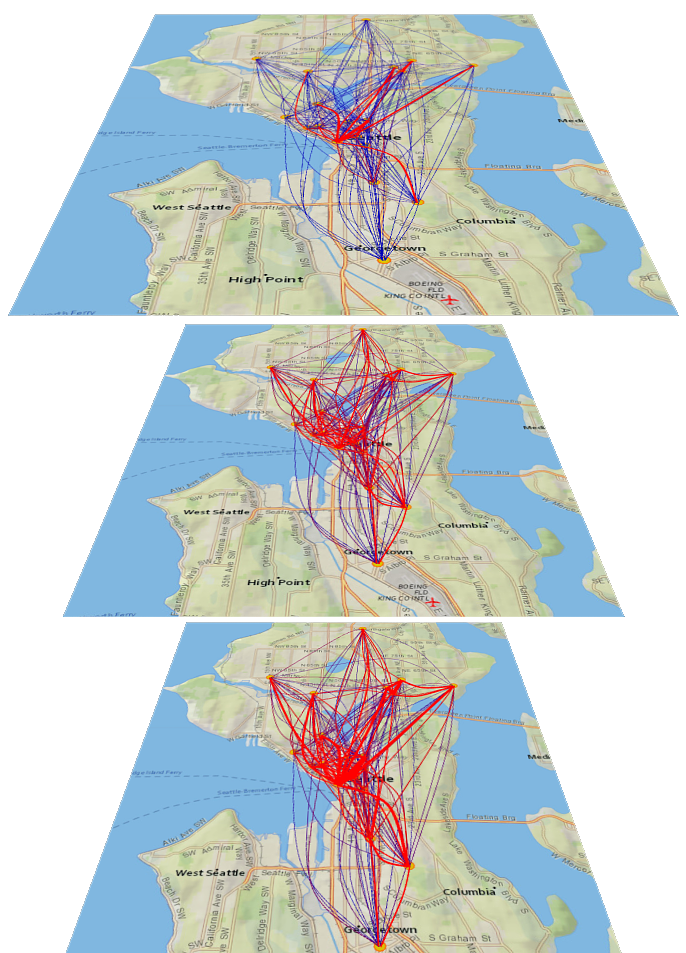}
         \caption{ $l_1$ norm}
         \label{fig:l1subfig}
     \end{subfigure}
    \hfill
    \begin{subfigure}[b]{0.49\textwidth}
         \centering
         \includegraphics[height=0.34\textheight,width=0.99\columnwidth]{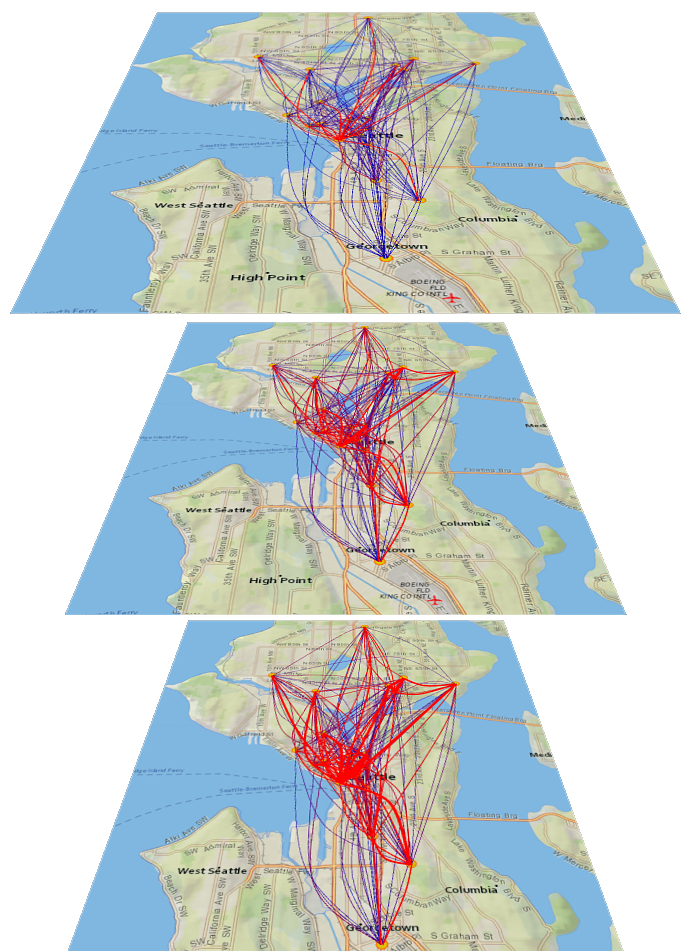}
         \caption{ $l_2$ norm}
         \label{fig:l2subfig}
     \end{subfigure}

%\begin{subfigure}{.45\textwidth}
%    \centering
%    \includegraphics[width=0.9\columnwidth]{pict/L1_Seattle_V2.jpg} 
%    \caption{  $l_1$-norm.} 
%    \label{fig:l1-solution}
%\end{subfigure}%
%\hfill
%\begin{subfigure}{.45\textwidth}
%    \centering
%    \includegraphics[width=0.9\columnwidth]{pict/L2_Seattle_V2.jpg} 
%    \caption{  $l_2$-norm.}
%    \label{fig:l2-solution}
%\end{subfigure}
\caption{  Comparison of the optimal prevention in the $20$-node model of the Seattle city, correspondent to Eq.~(\ref{eq:prevention-Q}) with $k=1$, computed for a particular instance of $J^{(0)}$ (leading to infection of the entire city, if not reduced) under $l_1$ (left sub-figure) and $l_2$ (right sub-figure). In each sub-graph initial instance, $J^{(0)}$, optimal correction,  $\widehat{J}^{(\text{\tiny corr})}$, and respective optimal reduction, $\widehat{J}^{(\text{\tiny corr})}-J^{(0)}$,  are shown at the bottom, middle and top layers respectively. Wider lines and red colors (in the red-to-dark blue color-coding) correspond to pair-wise links with higher values of $J$. (The Unfolding \cite{unfoldingSoftware} and Processing \cite{processingSoftware} software was used to create the Figure.)}
\label{fig:l1+l2-solutions}
\end{figure}

\begin{figure}[ht]
\centering
    \begin{subfigure}{0.3\textwidth}
        %\centering
        \includegraphics[height=0.2\textheight,width=0.99\columnwidth]{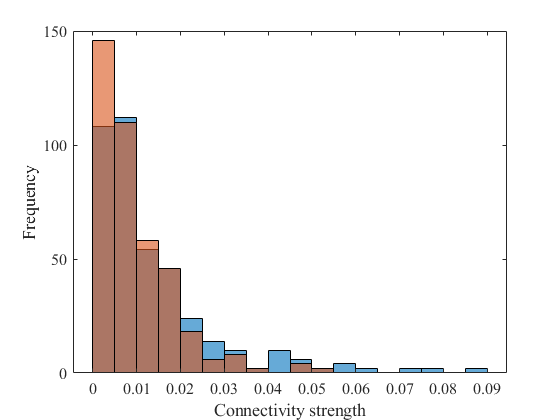}
        \caption{  Histogram  of the connectivity degree in the case of the $l_1$ norm.}
    \end{subfigure}
    \hspace{4.2cm}
    \begin{subfigure}{0.3\textwidth}
        %\centering
        \includegraphics[height=0.2\textheight,width=0.99\columnwidth]{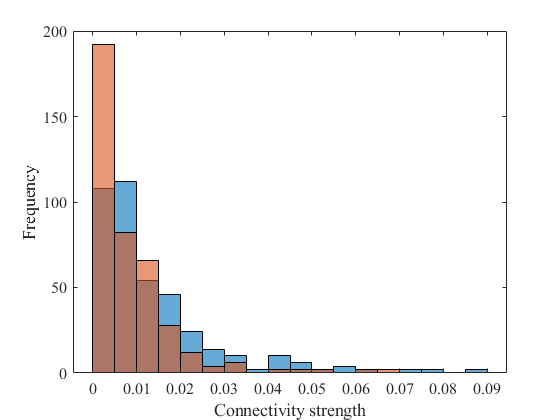}
        \caption{  Histogram  of the connectivity degree in the case of the $l_2$ norm.}
    \end{subfigure}
    \vfill
    \begin{subfigure}{0.45\textwidth}
        %\centering
        \includegraphics[height=0.28\textheight,width=0.99\columnwidth]{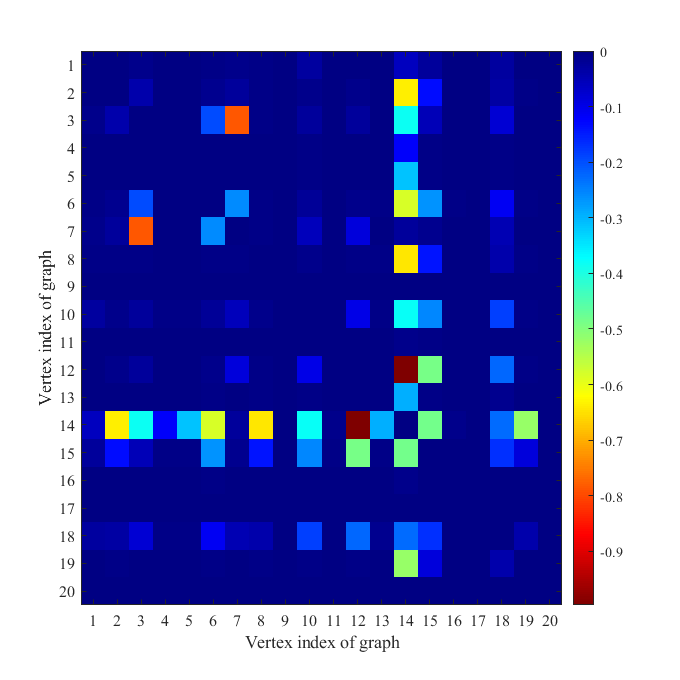}
        \caption{  Heat map of the reduction graph/matrix normalized by $J^{(0)}$ in the case of the $l_1$ norm, $\frac{(\widehat{J}^{(\text{\tiny corr})}-J^{(0)})}{max(J^{(0)})}$}
    \end{subfigure}
    \hfill
    \begin{subfigure}{0.45\textwidth}
        %\centering
        \includegraphics[height=0.28\textheight,width=0.99\columnwidth]{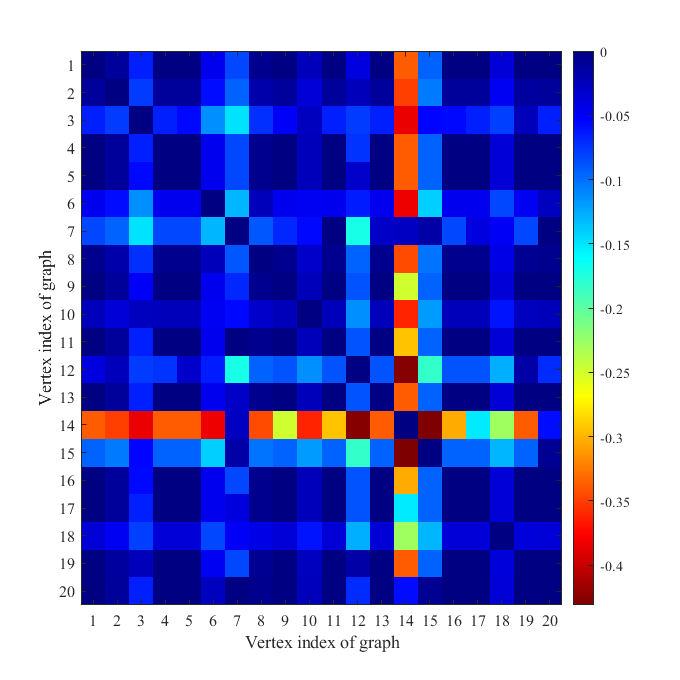}
        \caption{  Heat map of the reduction graph/matrix normalized by $J^{(0)}$ in the case of the $l_2$ norm, $\frac{(\widehat{J}^{(\text{\tiny corr})}-J^{(0)})}{max(J^{(0)})}$}
    \end{subfigure}
\caption{  Sub-figure (a) and Sub-figure (b),  correspondent to $l_1$ and $l_2$ cases respectively, present histogram  of the connectivity degree of the infected instance (blue) and corrected instance (orange) for the case of the Seattle city shown in Fig.~\ref{fig:l1+l2-solutions}. Sub-figure (c) and Sub-figure (d) show the heat-map representation of the reduction graphs normalized by $J^{(0)}$ from the middle layers in Fig.~\ref{fig:l1+l2-solutions}, where color coding of the $(i,j)$ element of the symmetric $20\times 20$ matrix is chosen proportional to $(\widehat{J}^{(\text{\tiny corr})}-J^{(0)})_{ij}$. (The Matlab \cite{MATLAB:2021} was used to create the figure.)}

%Comparison of connectivity revisions for the Seattle city under $l_1$ and $l_2$ norms. In sub-figures (a) and (b), the blue histogram shows the original connectivity and the orange histogram shows the optimal reduced connectivity. Sub-figures (c) and (d) show the connectivity reduction from the original state.}
\label{fig:l1l2Details}
\end{figure}

{  Our initial reason for choosing $l_1$ in the experiments discussed above was based on empirical expectation that $l_1$, as the closest of the convex options to $l_0$, will enforce sparsity of the optimal reduction in the pair-vise influences/traffic patterns. However, by choosing to work with the $l_1$-norm, we certainly did not mean to suggest that other norms cannot be used, especially if propounded by public health and/or government experts -- after all it is the professional judgment of these experts which is expected to drive practical choices of the cost function. Therefore, %(and inspired by respective remark of an antonymous referee questioning our initial choice of the $l_1$ norm), 
we also experimented with the $l_2$ norm and more generally with a convex mixture (interpolation between) $l_1$ and $l_2$ norms. These richer experiments are illustrated in Fig.~\ref{fig:l1+l2-solutions} and Fig.~\ref{fig:l1l2Details}. 

Fig.~\ref{fig:l1+l2-solutions} compares initial infectious patterns and optimal mitigation in the $l_1$ (left sub-figure) and $l_2$ (right sub-figure) cases. The experiments suggest (consistently with the original hypothesis) that significant reduction (of influence/traffic) in the case of $l_1$ norm is observed at a fewer pair-wise links and, most interestingly, these reductions are NOT all along the links which are most intense (traveled less) in the original instance. On the contrary, in the $l_2$ case there are more edges with comparable reductions and these reductions are largely across the links which show heavy influences/traffic already in the bare cases (without correction) which require prevention. 

Fig.~\ref{fig:l1l2Details} extends information provided in Fig.~\ref{fig:l1+l2-solutions}. Sub-figures (a) and (b) show histograms  of the connectivity degree of the infected instance (blue) and corrected instance (orange) for the use case of the Seattle city from Fig.~\ref{fig:l1+l2-solutions}. We observe that there are fewer nodes of high degree in the corrected graph. Sub-figure (c) and Sub-figure (d) show the heat-map representation of the reduction graphs from the middle layers in Fig.~\ref{fig:l1+l2-solutions}, where color coding of the $(i,j)$ element of the symmetric $20\times 20$ matrix is chosen proportional to $(\widehat{J}^{(\text{\tiny corr})}-J^{(0)})_{ij}$. Also, the histogram sub-figure show clearly that reduction mainly applies to immediate vicinity of the highest degree node in the downtown area of the Seattle city.}

\section*{Discussion}\label{sec:Conclusions}

In this manuscript, written specifically for an interdisciplinary audience of researchers in mathematical, physical and engineering sciences, we follow our prior work \cite{GM-Pandemic} (aimed primarily at experts in public health and epidemiology) and explain challenges in inference (prediction) and control (prevention) of pandemics. We use the language of GMs,  which is one powerful tool in the modern arsenal of Data Science and AI, and state the Prediction Challenge as a MAP optimization over an attractive Ising model, which can be expressed generically as a solution of a tractable Linear Programming (LP). We then turn to the analysis of the prevention problem, which one formulates if the aforementioned prediction solution suggests that the probability of significant infection is above a pre-defined (by the public health experts) tolerance threshold.  We show that in its simplest formulation, the prevention problem is equivalent to finding minimal $l_1$ projection to the safety polytope, where the latter is defined by solving the aforementioned prediction problem.  In general, the polytope does not allow a description non-exponential in the size of the system. However, we suggested an approximation that allows to approximate the safety polytope efficiently - that is, linearly in the number of the initial infection patterns. The approximation is justified (empirically,  with supporting theoretical arguments,  however not yet backed by a mathematically rigorous theory) in the case when the interaction graph of the system (e.g., related to the system/city transportation and human-to-human interaction network) is sufficiently dense. We conclude by providing a quasi-realistic experimental demonstration on the GM of Seattle. 

{  As a disclaimer but also a path forward, we would like to point out that this manuscript is not there yet where we would dream it to be in terms of practical use by heath-care authorities/practitioners.  However, the manuscript suggests an important, and yet missing in the literature, path towards the applications.  We focused in the manuscript on developing methodology which we intend to make useful for practical purposes in the future. This will require accurate calibration of the pairwise and singleton terms in the Ising (Graphical) models, e.g., as mentioned above, based on "learning" the rates from much better developed ABMs. When such calibration of the parameters in the Ising model is done, the pair-wise terms (J) and the singleton terms (h) may be interpreted as corresponding to pair-wise "traffic" or "influences" between aggregated areas and singleton biases, e.g., related to degree of vaccination or compliance with the local public health mandates. Then,  the outcome of this paper methodology will become practical, e.g., resulting in (optimal) recommendations of the traffic/influence and local modifications to avoid a disastrous spread of infection.}

We conclude with an incomplete list of technical challenges, presented in the order of importance (subjective), which need to be resolved to make the powerful GM approach to pandemic prediction and prevention practical: 
\begin{itemize}
    \item Build a hierarchy of {  Agent Based and} Probabilistic Graphical Models which allow more accurate (than Ising model) representation of the infection patterns over geographical and community graphs. {  (See also our remark above about making the modeling more realistic.)} The models may be both of the static (like Ising) or dynamic (like Independent Cascade Model) types. Extend the notion of the Safety Region (polytope) to the new GM of pandemics.
    
    \item Consider the case when the resolution of the Prediction Challenge problem returns a positive answer - most likely future state of the system is safe,  and then develop the methodology which allows estimating the probability of crossing the safety boundary. In other words, we envision formulating and solving in the context of the GM a problem which is akin to the one addressed in \cite{2019Owen}: estimate the probability of finding the system outside of the Safety Polytope. 
    
    \item Construct other (than two-mode) approximations to the Safety Polytope. In particular, approximations built on sampling of the boundaries of the safety polytope and learning (possibly reinforcement learning) are needed. 
    
    \item Develop the asymptotic (thermodynamic limit) theory which allows validating (and/or correcting systematically) the efficient (two-mode and other) approximations of the Safety Polytope. 
\end{itemize}

%\section*{Methods}

%Topical subheadings are allowed. Authors must ensure that their Methods section includes adequate experimental and characterization data necessary for others in the field to reproduce their work.

\bibliography{mitigation}

\section*{Acknowledgments}

This work was supported in part by NSF via \#2027072 "RAPID:Infer   and   Control   Global   Spread   of   Corona-Virus   with Graphical Models" project. The work of M.K. was supported by the Analytical center at Skoltech %under the RF Government 
(subsidy agreement 000000D730321P5Q0002, Grant No. 70-2021-00145 02.11.2021). The work of V.P. was partially supported by the Swedish Research Council and the Swedish Transport Administration.

\section*{Author contributions statement}

M.C. formulated the problem,  M.K. developed the two-mode approximation and conducted the projection experiments, A.M.E.S prepared Seattle data and conducted the experiments, A.E. and V.P. analyzed the results.  All authors contributed to writing of the manuscript. 

%\section*{Additional information}
%To include, in this order: \textbf{Accession codes} (where applicable); \textbf{Competing interests} (mandatory statement). 
%The corresponding author is responsible for submitting a \href{http://www.nature.com/srep/policies/index.html#competing}{competing interests statement} on behalf of all authors of the paper. This statement must be included in the submitted article file.

\end{document}